\documentclass{article}
\usepackage{amssymb}

\input{tcilatex}
\begin{document}

\date{\today}
\title{Maximum Force and Naked Singularities in Higher Dimensions}
\author{John D. Barrow \\
%EndAName
DAMTP, University of Cambridge, Wilberforce Rd.\\
Cambridge CB3 0WA, United Kingdom\\
email J.D.Barrow@damtp.cam.ac.uk}
\maketitle
\date{}

\begin{abstract}
We discuss the existence of maximum forces in $3+1$ dimensional spacetimes
and show that the existence of a mass-independent maximum force does not
occur in general relativity in spaces of more than three dimensions$.$
Instead, the maximum force increases with the masses of merging objects as $%
M^{\frac{N-3}{N-2}}$ and allows unbounded gravitational forces to occur.
This suggests that naked singularities can arise in more than three
dimensions because they are unprotected by a maximum force at the horizon
surface. This creates a new perspective on the expectation of naked
singularities in higher dimensions.
\end{abstract}

The concept and name of a 'Cosmic Censor' in physics first appears in
Medieval times \cite{grant, JDB}, when the key problem that exercised the
natural philosophers of the day was whether a physical vacuum could exist
locally in the universe. Aristotle had said 'no' because otherwise a body
could move unimpeded through it and achieve unlimited speed. St. Augustine
had said 'no' because the vacuum was the antithesis of God -- the state of
non-being that marked the greatest evil. But the issue was subjected to a
much more critical analysis in the 13th and 14th centuries by thinkers like
Roger Bacon, Walter Burley, and Blasius of Parma, who created a series of
thought experiments that attempted to show that an instantaneous vacuum
might be created by separating highly polished parallel metal sheets, or
instantaneously after dropping one of them to the ground, or allowing some
water to flow out of a clepsydra, or water catcher\footnote{%
The clepsydra was a metal vessel with holes in the base and a narrow neck.
When immersed in water with the neck open it would fill with water which
would flow out through the holes when lifted out of the water. But if a
finger sealed the neck while it was submerged the water would not flow out,
stopping the formation of a vacuum by not falling and opposing Aristotle's
physics.}. In each case a short-lived vacuum appeared to be formed
immediately after the plates separated or hit the ground, but arguments
raged over whether imperfections and the intervening air stopped this
happening in practice. Others were happy to allow an ephemeral vacuum to
form but not a stable one. All this is very reminiscent of the debates in
some areas of modern physics today about the formation of naked
singularities. A key point in the Medieval debate was Roger Bacon's
objection to the use of a negative principle to explain what we see in
Nature. Such principles were powerful vetoes but they still permitted many
things to occur that were not seen. Thus, the avoidance of a vacuum forming
in the clepsydra could also be explained by the walls imploding as well as
by the response of the water to pressure.~To resolve this ambiguity about
will actually happen, Burley introduced the concept of the 'Celestial Agent'
to stave off the possibility of a local vacuum forming. It could intervene
to explain why the clepsydra's water behaved 'unnaturally' by not falling
downwards rather than imploding the walls of the vessel to stop a vacuum
forming. This Celestial Agent was a distant forerunner of the Cosmic Censor,
that is (hopefully!) just a shorthand for as yet unknown properties of
general relativity. The modern Cosmic Censor \cite{penrose} forbids the
formation of local naked singularities: the Medieval Celestial Agent forbad
the formation of a local vacuum.

\ The modern study of naked singularities and cosmic censorship is an active
field that has discovered a significant dimensional dependence in the
problem. The situation in three space dimensions appears simpler and more
constrained than in higher dimensions where many variants of black holes and
singularities have been found \cite{ER}. Here, we present another
perspective on naked singularities in different dimensions that is
suggestive of why naked singularities spaces with more than three dimensions
might be expected. It is based on the possibility of a maximum force in
general relativity.

One of the features of Newtonian gravity under a Newtonian inverse-square
law is that the force between point masses can become unboundedly large as
there separation shrinks to zero. In general relativity, it appears that
this cannot happen because of the existence of black hole horizons. Once two
approaching particles of mass $M$ \ fall within their Schwarzschild radius
an event horizon appears around them and the maximum force between them is
given by the Newtonian inverse-square force when their individual horizons
touch. This leads to a maximum force of $c^{4}/4G.$ This is also the maximum
tension that a cosmic string can possess and arises when the defect angle is
equal to $2\pi $ radians. This maximum force has a special fundamental
classical significance because it is the Planck unit of force but does not
contain Planck's constant, $h$. This feature is shared by the Planck units
for velocity, $c,$ magnetic moment to angular momentum ratio $(G^{1/2}/c)$,
power $\left( c^{5}/4G\right) $, mass flow rate $(c^{3}/4G)$, and luminosity 
$\left( c^{5}/4G,\right) ,$ \cite{JB,JB2}, all of which appear to signal the
existence of physical maxima. These purely non-quantum Planck units always
signal something fundamental in physics. In this essay we suggest that the
maximum force condition is telling us something about naked singularities in
higher dimensions.

First, we look at the $N$-dimensional analogue of the simple $N=3$
derivation of a maximum force between two black holes of mass $M$ separated
by the horizon distance. For the Newtonian $N$-dimensional black hole we
follow Michell \cite{mich} and Laplace \cite{lap,lap2} in the
three-dimensional case and look for objects with escape velocity equal to
the speed of light. This condition for escape speed equals light speed, $c$,
is

\[
\frac{c^{2}}{2}=\Phi _{N}, 
\]%
where the $N$-dimensional gravitational force $F_{N}$ and potential $\Phi
_{N}$ are ($N\neq 2$)

\begin{equation}
F_{N}=\frac{G_{N}M^{2}}{r^{N-1}},\text{ \ }\Phi _{N}=-\frac{(N-2)G_{N}M}{%
r^{N-2}}.  \label{a}
\end{equation}

So the Newtonian $N$-dimensional black hole has horizon radius

\begin{equation}
R=R_{g}=\left( \frac{2GM(N-2)}{c^{2}}\right) ^{1/(N-2)}.  \label{b}
\end{equation}

The force between two black holes of mass $M$ separated by $R_{g}$ is

\begin{equation}
F_{N}=\frac{G_{N}M^{2}\ }{R_{g}^{N-1}}.  \label{c}
\end{equation}

This is

\begin{equation}
F_{N}=\left[ 2(N-2)\right] ^{\frac{1-N}{N-2}}G_{N}^{\frac{-1}{N-2}}M^{\frac{%
N-3}{N-2}}c^{\frac{2(N-1)}{N-2}}.  \label{1}
\end{equation}%
This is the candidate for the maximum force. As a check, we note that $F_{3}$
reduces to $c^{4}/4G$ $=3.3\times 10^{43}N$, and there is no dependence on $%
M $. This is significant. Note that $R_{g}$ vanishes when $N=2$ in which
case here are no gravitational forces as the Weyl tensor vanishes and the
quantity $GM$ \ has no dimensions, so when $N=2$ there is no location for a
horizon.

In $N$-dimensional space, the Planck unit of force is%
\begin{equation}
F_{pl}=G^{2/(1-N)}c^{(5+N)/(N-1)}h^{(3-N)/(1-N)}.  \label{d}
\end{equation}%
We see that this is only independent of $h$ and equal to $F_{N\text{ }}$when 
$N=3$. In other dimensions, it is partly quantum because $h$ features
explicitly.

Previously, we found that $MA^{N-2}$ was the classical ($h$-independent)
Planck unit in $N$ dimensions, where $A$ is the Planck unit of acceleration,
and so it is only a force, $MA$, when $N=3.$In general, we have:

\begin{equation}
MA^{N-2}=\frac{(F_{N})^{N-2}}{M^{N-3}}=\frac{(MA)^{N-2}}{M^{N-3}}=\left[
2(N-2)\right] ^{1-N}G_{{}}^{-1\ }\ c^{2(N-1)}.  \label{e}
\end{equation}

This argument applies in $N$-dimensional general relativity also. The $N$%
-dimensional generalisation of the Schwarzschild vacuum solution was found
by Tangherlini \cite{tang} in 1963. The metric is given by

\begin{equation}
ds^{2}=-c^{2}(1-\frac{R_{g}^{N-2}}{r^{N-2}})dt^{2}+\frac{dr^{2}}{(1-\frac{%
R_{g}^{N-2}}{r^{N-2}})}+r^{2}d\Omega _{N-1}^{2},  \label{f}
\end{equation}

where the metric of an $(N-1)$-dimensional unit sphere is given by 
\begin{equation}
d\Omega _{N-1}^{2}=d\theta _{1}^{2}+\sin ^{2}\theta _{1}d\theta _{2}^{2}+...+%
\text{ }(\sin ^{2}\theta _{1}...\sin ^{2}\theta _{N-2})d\theta _{N-1}^{2}.
\label{g}
\end{equation}

The horizon radius is located at the radius

\begin{equation}
r=R_{g}=\ \left[ \frac{16\pi GM}{(N-1)c^{2}\Omega _{N-1}}\right] ^{\frac{1}{%
N-2}},  \label{i}
\end{equation}%
where $\Omega _{N-1}$ is the area of a $N-1$ sphere with unit radius:%
\begin{equation}
\Omega _{N-1}=2\pi ^{N/2}/\Gamma (\frac{N}{2}).  \label{h}
\end{equation}

So, for $N=3$, $\Gamma (3/2)=\pi ^{1/2}/2,$ $\Omega _{2}=4\pi $ and $R_{g}=\
2GM/c^{2}$, as expected$.$

The metric \ref{f} provides a vacuum solution of the $N$-dimensional
Einstein theory, describing static, asymptotically-flat higher-dimensional
black holes. The black hole horizon exists at $R_{g}$.

Now, to find the maximum force allow two $N$-dimensional black holes of mass 
$M$ to touch horizons and calculate the force between them using the force
law in $N$ dimensions (the same maximum force is found if the masses are
unequal\footnote{%
Use the inequality $(\sqrt{M_{1}-\sqrt{M_{2}})^{2}}=M_{1}+M_{2}-2\sqrt{%
M_{1}M_{2}}\geq 0.$%
\par
Hence, $M_{1}M_{2}\leq \frac{1}{4}(M_{1}+M_{2})^{2}$ in the formula for $%
F_{N}$ and we find for the maximum force $F_{N}(M_{1},M_{2})\leq
F_{N}(M_{1},M_{1}).$}):

Using an approach distance $r=2R_{g}$, we have

\begin{equation}
F_{N}=\ \frac{(N-2)8\pi GM^{2}}{(N-1)\Omega _{N-1}(2GM/c^{2})^{\frac{N-1}{N-2%
}}}\varpropto G^{\frac{-1}{N-2}}M^{\frac{N-3}{N-2}}c^{\frac{2(N-1)}{N-2\ }}.
\label{k}
\end{equation}
So, when $N=3$

\begin{equation}
F_{3}=\frac{4\pi GM^{2}c^{4}}{\Omega _{2}\mu ^{2}}\ =\frac{c^{4}}{4G}.
\label{l}
\end{equation}

Therefore, we can have arbitrarily large forces as we increase the mass, $M$%
, when $N>3,$ since

\begin{equation}
F_{N}\varpropto M^{\frac{N-3}{N-2}},  \label{m}
\end{equation}

and the black hole horizon is no longer a protection against physical forces
of arbitrarily large magnitude. Such a protection only happens in
three-dimensional space. In other dimensions, when $N>3$ this force can be
arbitrarily large as $M$ increases.

The maximum force occurs when the event horizon forms around the touching
masses when their separation equals the black hole horizon size. A horizon
appears if you try to exceed the maximum force. The horizon prevents any
physical configuration that could surpass the maximum force limit. So, we
see that in three dimensions the horizon is a mass-independent barrier to
all attempts to exceed the maximum force limit. But in higher dimensions
there is not a horizon barrier and with increasing mass of the merging
objects an arbitrarily large force can occur without the event horizon
stopping it.

\ \ This is very suggestive that in three dimensions the avoidance of
arbitrarily large forces and consequent naked singularities is expected, but
in higher dimensions ($N>3$) the possibility is open to create arbitrarily
large forces and there is no obstruction to the creation of a naked
singularity, even if they are characterised by infinite gravitational
forces. In the presence of rotation we know that the situation becomes much
more baroque, with a plethora of new possibilities -- black rings, black
saturns and strings \cite{ER} -- as well as the more conventional
Myers-Perry \cite{MP} generalisations of the Kerr metric. The extension of
our analysis to rotating lack holes is a problem for the future. The physics
of higher-dimensional black holes is much richer than in three-dimensional
space because the number of rotational axes is no longer equal to the number
of dimensions and the balance between centrifugal forces, $J^{2}/M^{2}r^{2},$
which are dimension independent and gravitational force, which depends on
dimension via eq. (\ref{a}), changes when $N=3,4,$and $N\geq 5.$

In conclusion, we see that, in contrast to the situation when $N=3$, a
mass-independent maximum force does not exist in general relativity when $N>3
$ and this suggest that naked singularities are more likely to form from
force strengths that are unprotected by a horizon in more than three
dimensions than in three-dimensional spaces. This offers a new perspective
on cosmic censorship.

Acknowldegements; I would like to thank Gary Gibbons for many discussions.
The author is supported by the STFC of the UK.

\end{document}